\documentclass[preprint,showpacs,preprintnumbers,amsmath,amssymb]{revtex4}

\usepackage{graphicx}
\usepackage{dcolumn}
\usepackage{bm}
\usepackage{hyperref}
\usepackage{latexsym}
\usepackage{color}
\begin{document}

\title{Eradicating Catastrophic Collapse in Interdependent Networks via Reinforced Nodes }

\author{Xin Yuan,$^2$ Yanqing Hu,$^{1,3,4}$
H. Eugene Stanley, $^2$  and Shlomo Havlin,$^5$}

\affiliation{$^1$School of Data and Computer Science, Sun 
Yat-sen University, Guangzhou 510006, China\\
$^2$Center for Polymer Studies and Department of
Physics, Boston University, Boston, Massachusetts 02215, USA\\
$^3$School of Mathematics, Southwest Jiaotong University, Chengdu 610031, China\\
$^4$Big Data Research Center, University of Electronic Science and Technology of China, Chengdu
611731, China\\
$^5$Department of Physics, Bar-Ilan University,  Ramat-Gan 52900,
Israel}

\date{12 May 2016}

\date{\today}


\begin{abstract}

In interdependent networks, it is usually assumed, based on percolation theory, that nodes become
nonfunctional if they lose connection to the network giant component. However, in reality,
some nodes, equipped with alternative resources, together with their
connected neighbors can still be functioning once disconnected from the
giant component. Here we propose and study a generalized percolation model that introduces a fraction of reinforced nodes in the
interdependent networks that can function and support their
neighborhood. We analyze, both analytically and via simulations, the order parameter$-$the functioning
component$-$comprising both the giant component and smaller components
that include at least one reinforced node. Remarkably, we find that for
interdependent networks, we need to reinforce only a small fraction of
nodes to prevent abrupt catastrophic collapses. Moreover, we find that the
universal upper bound of this fraction is 0.1756 for two interdependent
Erd\H{o}s-R\'{e}nyi (ER) networks, regular-random (RR) networks and
scale-free (SF) networks with large average degrees.  We also generalize
our theory to interdependent networks of networks (NON).  Our findings might yield insight for designing resilient interdependent infrastructure networks.

\end{abstract}

\maketitle


Complex networks often interact and depend on each other to function
properly
\cite{rinaldi2001identifying,little2002controlling,rosato2008modelling,buldyrev2010catastrophic,bashan2012network,
gao2012networks,helbing2013globally,jp2007structure}.
Due to interdependencies, these interacting networks may easily suffer
abrupt failures and face catastrophic consequences, such as the
blackouts of Italy in 2003 and North America in 2008
\cite{rosato2008modelling,buldyrev2010catastrophic,gao2012networks}. Thus,
a major open challenge arises as how to tackle the vulnerability of
interdependent networks. Virtually many existing theories
on the resilience of interacting networks have centered on the formation
of the largest cluster (called the giant component) \cite{buldyrev2010catastrophic,gao2012networks,coniglio1982cluster,Radicchi2015NP,Hu2014NP,boccaletti2014structure,
cohen2000resilience,newman2002spread,cohen2010complex}, 
and consider only the nodes in the giant component as functional, 
since all the small clusters do not have a connection to the majority of 
nodes, which are in the giant component.

However, in many realistic networks, in case of network
component failures, some nodes (which we call here reinforced nodes),
and even clusters containing reinforced nodes outside of the giant
component, can resort to contingency mechanisms or back-up facilities to
keep themselves functioning normally \cite{jenkins1995embedded,pepermans2005distributed,alanne2006distributed}. 
For example, small neighborhoods in a city when facing a sudden power 
outage could employ alternative facilities to sustain themselves. Consider 
also the case where some important internet ports, after their fiber links are 
cut off from the giant component, could use satellites \cite{henderson1999transport} 
or high-altitude platforms \cite{mohammed2011role} to exchange vital
information.  These possibilities strongly motivate us to generalize the 
percolation theory \cite{coniglio1982cluster,vicsek1994fractals}  to include a fraction of reinforced nodes that are capable 
of securing the functioning of the finite clusters in which they are located.  
We apply this framework to study a system of interdependent networks and 
find that a small fraction of reinforced nodes can avoid the catastrophic abrupt collapse.
 
In this paper we develop a mathematical framework based on percolation \cite{buldyrev2010catastrophic,gao2012networks,boccaletti2014structure,cohen2000resilience,brown2013monte} for studying
interdependent networks with reinforced nodes and find exact solutions
to the minimal fraction of reinforced nodes needed to eradicate
catastrophic collapses. In particular we apply our framework to study
and compare three types of random networks, (i) ER networks with a
Poisson degree distribution ($P(k)=e^{-\left\langle k
  \right\rangle}{\left\langle k \right\rangle}^k/k!$)
\cite{bollobas2001random}, (ii) SF networks with a power law degree
distribution ($P(k) \sim k^{-\lambda}$) \cite{albert2002statistical},
and (iii) RR networks with a Kronecker delta degree distribution
($P(k)=\delta_{k,k_0}$).  We find the universal upper bound for this
minimal fraction to be 0.1756 for two interdependent ER networks with
any average degree and SF and RR networks with a large average degree.

\section{Model}
\begin{figure}[h!]
\centering
   \includegraphics[width=0.4\textwidth, angle = 0]
 {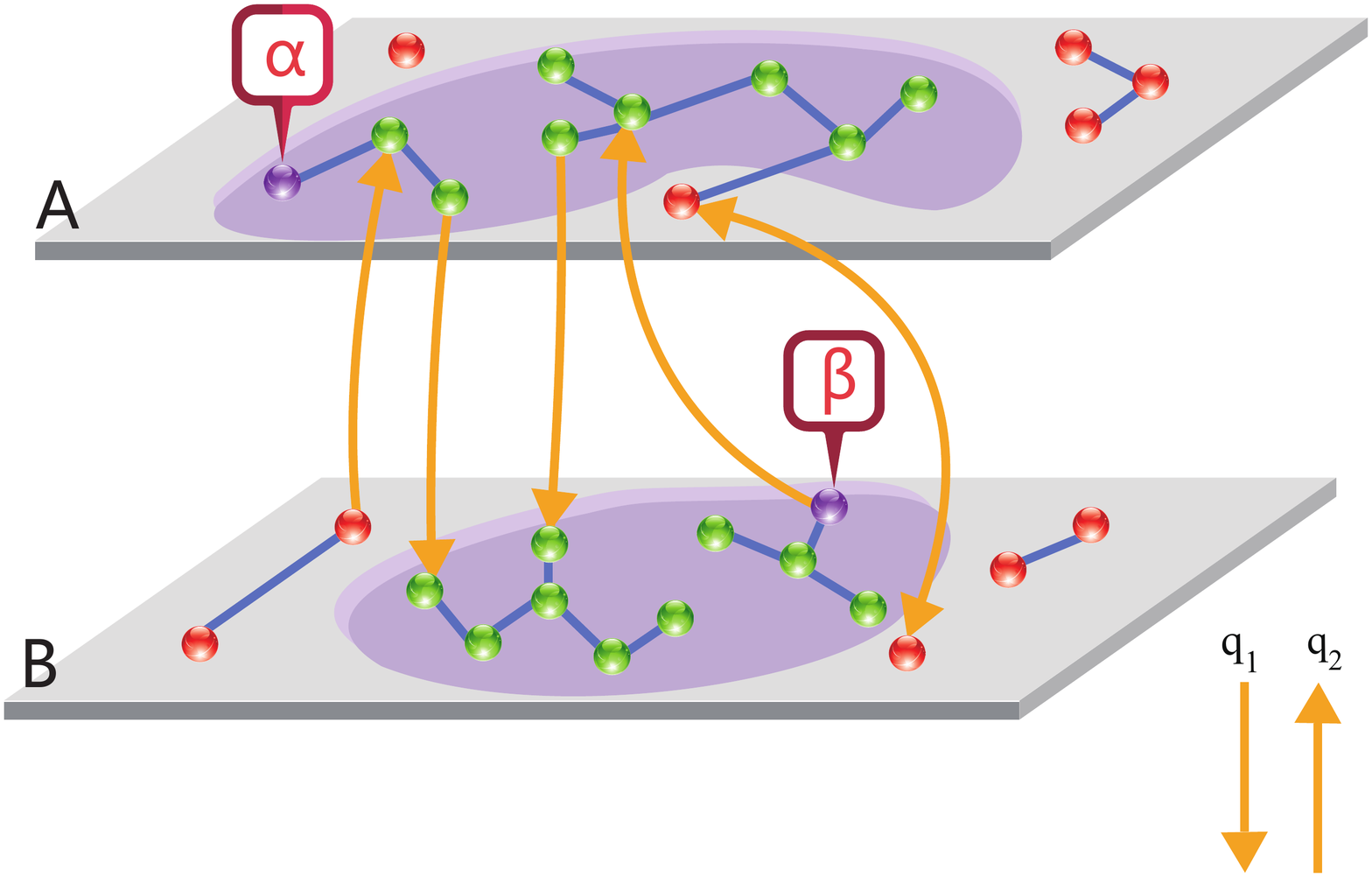}
\caption{Demonstration of the model studied
  here where two interdependent networks $A$ and $B$ have gone through
  cascading failures and reached a steady state. The yellow arrows
  represent a fraction $q_{A(B)}$ of nodes from network $A(B)$ depending
  on nodes from network $B(A)$ for critical support.  Reinforced nodes
  $\alpha$ and $\beta$ (purple circles) are nodes that survive and also
  support their clusters even if the clusters are not connected to the
  largest component. Some regular nodes (green circles) survive the
  cascading failures whereas some other regular nodes (red circles)
  fail. Note that the clusters of circles in the shaded purple areas
  constitute the functioning component studied in our model.}\label{Fig1} 
\end{figure}

Formally, for simplicity and without loss of generality, our model
consists of two networks $A$ and $B$ with $N$ nodes in each network (see
Fig.~\ref{Fig1}).  Within network $A$ the nodes are randomly connected
by $A$ links with degree distribution $P_A(k)$, while in network $B$ the
nodes are randomly connected by $B$ links with degree distribution
$P_B(k)$.  In addition, a fraction $q_A$ of nodes in $A$ are randomly
dependent (through dependency links) on nodes in network $B$ and a
fraction $q_B$ of nodes in network $B$ are randomly dependent on nodes
in network $A$ \cite{parshani2010interdependent}. We also assume that a
node from one network depends on no more than one node from the other
network and if a node $i$ in network $A$ is dependent on a node $j$ in
network $B$ and $j$ depends on a node $l$ in network $A$, then $l=i$ (a
no-feedback condition \cite{buldyrev2010catastrophic,gao2012networks,Hu2011PRE,hu2013PRE}).  
We denote $\rho_A$ and $\rho_B$ as the fractions of
nodes that are randomly chosen as reinforced nodes in network $A$ and
network $B$, respectively. In each network, together with the giant
component, those smaller clusters containing at least one reinforced
node make up the functioning component, as shown in
Fig.~$\ref{Fig1}$. The failure process is initiated by removing randomly
a fraction $1-p$ of nodes from each network. Therefore when nodes from
one network fail their dependent counterparts from the other network
must also fail.  In this case, an autonomous node (a node that does not
need support from the other network) \cite{parshani2010interdependent}
survives if it is connected to a functioning component of its own
network; a dependent node $n_0$ survives if both $n_0$ and the node it
depends on are connected to their own networks' functioning components.

We introduce the generating function of the degree distribution
$G_{A0}(x)=\sum_k P_A(k)x^k$ and the associated branching processes
$G_{A1}(x)=G^{'}_{A0}(x)/G^{'}_{A0}(1)$ \cite{newman2002spread}; similar
equations exist to describe network $B$.  At the steady state, using the
probabilistic framework
\cite{son2012percolation,baxter2010bootstrap,baxter2014weak,min2014multiple,feng2015simplified,min2015link,Bianconi2014PRE},
we denote $x$ ($y$) as the probability that a randomly chosen link in
network $A$ ($B$) reaches the functioning component of network $A$ ($B$)
at one of its nodes. Thus $x$ and $y$ satisfy the following
self-consistent equations (see \textit{SI Appendix}, section 2),
\begin{equation}\label{system1x}
 x=p\left[1-(1-\rho_A)G_{A1}(1-x)\right] \times 
 \left\lbrace1-q_A+pq_A\left[1-(1-\rho_B)G_{B0}(1-y)\right]\right\rbrace,
\end{equation}
and
\begin{equation}\label{system1y}
 y=p\left[1-(1-\rho_B)G_{B1}(1-y)\right]\times 
 \left\lbrace1-q_B+pq_B\left[1-(1-\rho_A)G_{A0}(1-x)\right]\right\rbrace. 
 \end{equation}
These two equations can be transformed into $x=F_1(p,y)$ and
$y=F_2(p,x)$, which can be solved numerically by iteration with the
proper initial values of $x$ and $y$.

Accordingly, the sizes of the functioning components are determined by
(see \textit{SI Appendix}, section 2)
\begin{equation}\label{system1mu}
P^{A}_{\infty}=p[1-(1-\rho_A)G_{A0}(1-x)]\times
\left\lbrace1-q_A+pq_A[1-(1-\rho_B)G_{B0}(1-y)]\right\rbrace,
\end{equation}
and
\begin{equation}\label{system2mu}
 P^B_{\infty}=p[1-(1-\rho_B)G_{B0}(1-y)]\times
 \left\lbrace1-q_B+pq_B[1-(1-\rho_A)G_{A0}(1-x)]\right\rbrace.
\end{equation}
If the system has an abrupt phase transition at $p=p_c^I$, the functions
$x=F_1(p,y)$ and $y=F_2(p,x)$ satisfy the condition
\begin{equation}\label{1storder}
\frac{\partial F_1(p_c^I,y^I)}{\partial y^I} \cdot\frac{\partial F_2(p_c^I,x^I)}{\partial x^I}=1, 
\end{equation}
namely, the curves $x=F_1(p_c^I,y)$ and $y=F_2(p_c^I,x)$ touch each
other tangentially at $(x^I,y^I)$
\cite{feng2015simplified,gao2013percolation}.

\section{Results}
\begin{figure}[h!]
 \centering
   \includegraphics[width=0.45\textwidth, angle = 0]{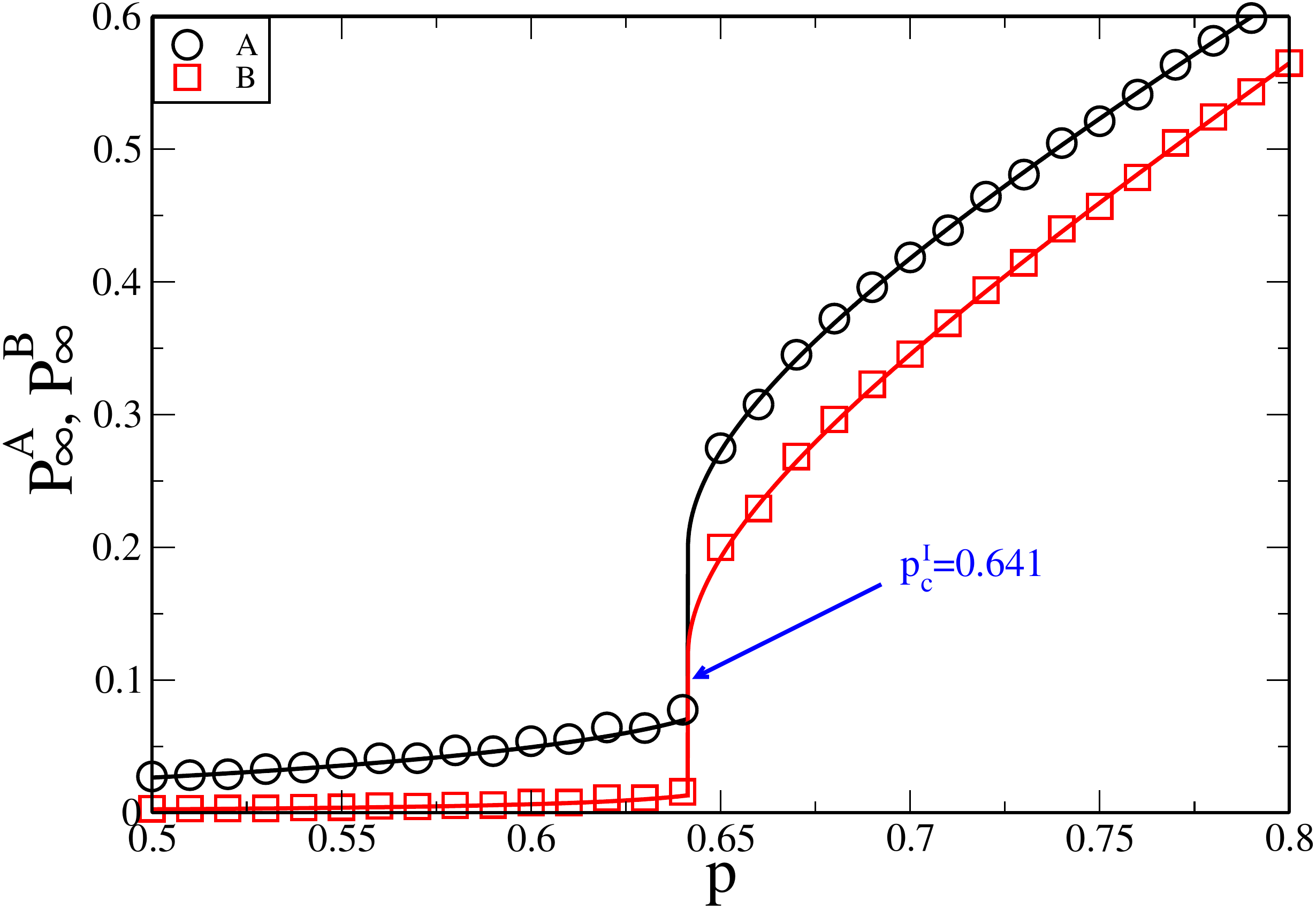}
\caption{ The sizes of functioning components
  as a function of $p$ for ER networks with $\rho_A=0.05$,
  $\rho_B=0.03$, $q_A=0.65$, $q_B=0.95$, $\left\langle k
  \right\rangle_A=4$, and $\left\langle k \right\rangle_B=5$. The
  simulation results (symbols) are obtained from two networks of $10^5$
  nodes and are in good agreement with the theoretical results (solid
  lines), Eqs.~\ref{system1mu} and \ref{system2mu}.  Note that for
  $\rho_A\neq 0$ and $\rho_B\neq 0$, network $A (B)$ always has at least
  a fraction $p^2 \rho_A\rho_Bq_A$ ($p^2 \rho_A\rho_Bq_B$) of nodes
  functioning after a fraction $1-p$ of nodes are removed from both
  networks. }\label{Fig2}
\end{figure}

For a general system of interdependent networks $A$ and $B$, $P^A_{\infty}$, $P^B_{\infty}$ and
the existence of $p_c^I$ can be easily determined numerically, using
Eqs.~\ref{system1x}-\ref{1storder}. As an example, Fig.~\ref{Fig2}
shows the excellent agreement between simulation and theory.

However it is important to find analytic expressions for $P^A_{\infty}$,
$P^B_{\infty}$ and $p_c^I$, at least for simpler cases, that can serve
as a benchmark to better understand simulated solutions of more
realistic cases. Thus here, for simplicity, we consider the symmetric
case where $P_A(k)=P_B(k)$, $\rho_A=\rho_B=\rho$ and $q_A=q_B=q$.  This
symmetry readily implies that $x=y \equiv F(p,x)$, reducing
Eqs.~\ref{system1x} and \ref{system1y} to a single
equation. Similarly, it renders $P^A_{\infty}=P^B_{\infty}\equiv
P_{\infty}$ and transforms Eq.~\ref{1storder} to $\frac{\partial
  F(p_c^I,x^I)}{\partial x^I} \cdot\frac{dx^I}{dx^I}=1$, i.e.,
$\frac{\partial F(p_c^I,x^I)}{\partial x^I}=1$.  Using
Eqs.~\ref{system1x}-\ref{1storder}, we derive $p_c^I$ and
$P_{\infty}$ rigorously (see \textit{SI Appendix}, section 3).

\begin{figure}[h!]
\centering
   \includegraphics[width=0.45\textwidth, angle = 0]
   {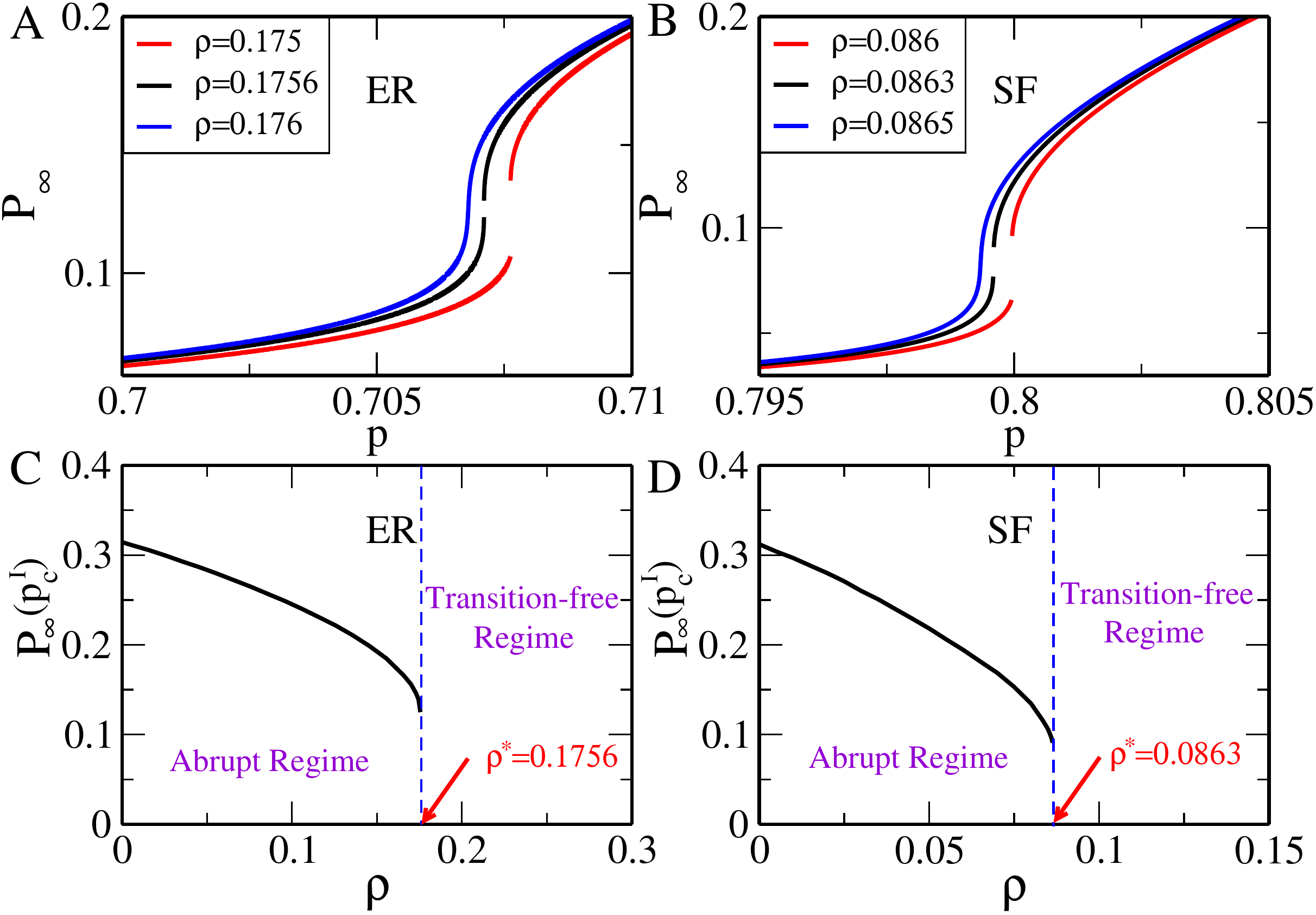}
\caption{ Percolation properties of symmetric
  interdependent ER and SF networks.  (\textit{A}),(\textit{B}) Demonstration of the
  behavior of $P_\infty(p)$ around $\rho^{\ast}$ for (\textit{A}) ER network with
  $\left\langle k \right\rangle =4$, $q=1$ and (\textit{B}) SF networks with
  $P(k)\sim k^{-\lambda}$, $\lambda=2.7$, $k_{\rm min}=2$, $k_{\rm max}=2048$
  and $q=1$.  (\textit{C}),(\textit{D}) The size of the functioning component
  $P_\infty(p_c^I)$ at the abrupt collapse as a function of $\rho$. We
  find $\rho^{\ast}$ for both cases as highlighted in the graphs.  }\label{Fig3}
\end{figure}

Surprisingly, we find that even for a system built with a relatively
high dependency coupling there exists a specific value $\rho^{\ast}$
that divides the phase diagram into two regimes.  Specifically, if $\rho
\leq \rho^{\ast}$, the system is subject to abrupt transitions; however,
if $\rho >\rho^{\ast}$, the abrupt percolation transition is absent in
the system because the giant component changes from a first-order phase
transition behavior to a second-order phase transition behavior (see
\textit{SI Appendix}, section 3). Therefore $\rho^{\ast}$ is the minimum
fraction of nodes in each network that need to be reinforced in order to
make the interdependent system less risky and free from abrupt
transitions. Moreover, $\rho^{\ast}$ satisfies the condition (see
\textit{SI Appendix}, section 3)
\begin{equation}
 \frac{dp_c^{I}}{dx^I}\vert_{\rho=\rho^{\ast}}=0.
\end{equation} 
Figure~\ref{Fig3} shows the existence of $\rho^{\ast}$ for systems of
fully interdependent ER networks ($\rho^{\ast}\approx0.1756$) and
scale-free (SF) networks ($\rho^{\ast}\approx0.0863$), respectively;
Figs.~\ref{Fig3}\textit{A} and \ref{Fig3}\textit{B} depict the dramatic 
behavior change of the functioning components as $\rho$ increases 
slightly from under $\rho^{\ast}$ to above $\rho^{\ast}$.

We next solve this critical value $\rho^{\ast}$ as a function of $q$ and
$\left\langle k \right\rangle$ for two interdependent ER networks as
(see \textit{SI Appendix}, section 3.1),
\begin{equation}\label{u_Arho}
 \rho^{\ast}=1-\frac{exp\left\lbrace\frac{1}{2}\left[1-{\left\langle k \right\rangle (1-q)^2}/{2q}\right]\right\rbrace}{2-\sqrt{{\left\langle k \right\rangle (1-q)^2}/{2q}}} ,
\end{equation}
where $q_0\leq q \leq 1 $ and $q_0$ is the minimum strength of
interdependence required to abruptly collapse the system
\cite{gao2011robustness}. If we set $\rho^{\ast}=0$ in
Eq.~\ref{u_Arho}, $q_0$ can be obtained from ${\left\langle k
  \right\rangle (1-q_0)^2}/{2q_0}=1$ as $ q_0=\left(1+{\left\langle k
  \right\rangle }-\sqrt{2\left\langle k \right\rangle
  +1}\right)/{\left\langle k \right\rangle}$, as found in
Refs.~\cite{gao2013percolation,parshani2011critical}. Applying Taylor
expansion to Eq.~\ref{u_Arho} for $q\rightarrow q_0$, we get the
critical exponent $\beta_1$, defined via $\rho^{\ast}\thicksim
(q-q_0)^{\beta_1}$ with $\beta_1=3$.

Hence for any $q \in [q_0,1]$ we first calculate $\rho^{\ast}$ using
Eq.~\ref{u_Arho} then $p_c^I$ corresponding to this $q$ and 
$\rho^{\ast}$ can be computed as (see \textit{SI Appendix}, section 3.1)
\begin{equation}
 p_c^{I}(q, \rho^{\ast})=\left[2-(1-q)\sqrt{{\left\langle k
       \right\rangle }/{2q}}\right]/{\sqrt{2\left\langle k \right\rangle
     q}},\label{p_c} 
\end{equation} 
and the size of the functioning component at this $p_c^{I}$ is
\begin{equation}
P_{\infty}(p_c^I)=[1-{\left\langle k \right\rangle
    (1-q)^2}/{2q}]/{2\left\langle k \right\rangle }.\label{mu} 
\end{equation}
The behavior of the order parameter $P_{\infty}(p)$ near the critical
point is defined by the critical exponent $\beta_2$, where
$P_{\infty}(p\rightarrow p_c^I) \thicksim (p-p_c^I)^{\beta_2}$ with
$\beta_2=1/3$ if $\rho=\rho^{\ast}$ and $\beta_2=1/2$ otherwise (see
\textit{SI, Appendix}, section 3.1.1 and
Ref. \cite{parshani2010interdependent}). Similar scaling behaviors have
been reported in a bootstrap percolation problem
\cite{baxter2010bootstrap}.

\begin{figure}[h!]
\centering
   \includegraphics[width=0.45\textwidth, angle = 0]
   {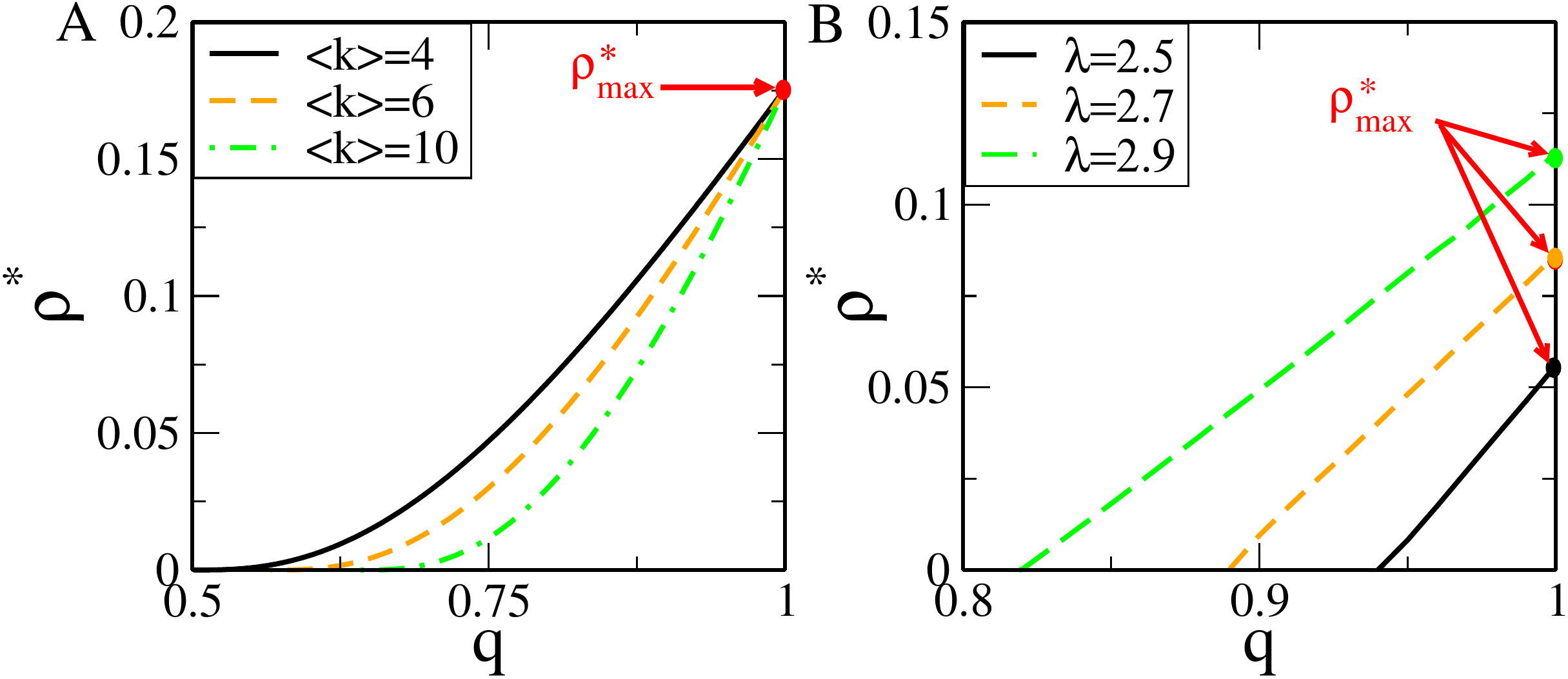}
\caption{ (\textit{A}) $\rho^{\ast}$ as a function of
  $q$ for symmetric ER networks with different values of $\left\langle k
  \right\rangle$. The results are obtained using Eq.~\ref{u_Arho} and
  these curves converge at the point $(1,0.1756)$. (\textit{B}) $\rho^{\ast}$ as
  a function of $q$ for symmetric SF networks with $k_{\rm min}=2$ and
  different values of $\lambda$. The results are obtained from numerical
  calculations (Eq.~(30) in \textit{SI Appendix}, section 3. We always have
  $\rho^{\ast}_{\rm max}$ at $q=1$ corresponding to the fully interdependent
  scenario.}\label{Fig4} 
\end{figure}

In Fig.~\ref{Fig4}\textit{A} we plot $\rho^{\ast}$ from Eq.~\ref{u_Arho} as a
function $q$ for several different values of $\left\langle k
\right\rangle$. Interestingly, at $q=1$, namely, for two fully
interdependent ER networks, we find, for all mean degrees, the maximum
of $\rho^{\ast}$ to be
\begin{equation}
\rho^{\ast}_{\rm max}=1-e^{{1}/{2}}/2\approx0.1756, 
\end{equation}
which is independent of $\left\langle k \right\rangle$.  In
Fig.~\ref{Fig4}\textit{B} we plot $\rho^{\ast}$ as a function of $q$ for
several degree exponents $\lambda$ of SF networks.  Here $\rho^{\ast}$
increases as $\lambda$ increases and takes its maximum
$\rho^{\ast}_{\rm max}$ at $q=1$, corresponding to the fully interdependent
case, which is the most vulnerable.  Thus if the dependency strength $q$
is unknown, $\rho^{\ast}_{\rm max}$ is the minimal fraction of reinforced
nodes, that can prevent catastrophic collapse.

\begin{figure}[h!]
\centering
   \includegraphics[width=0.45\textwidth, angle = 0]
   {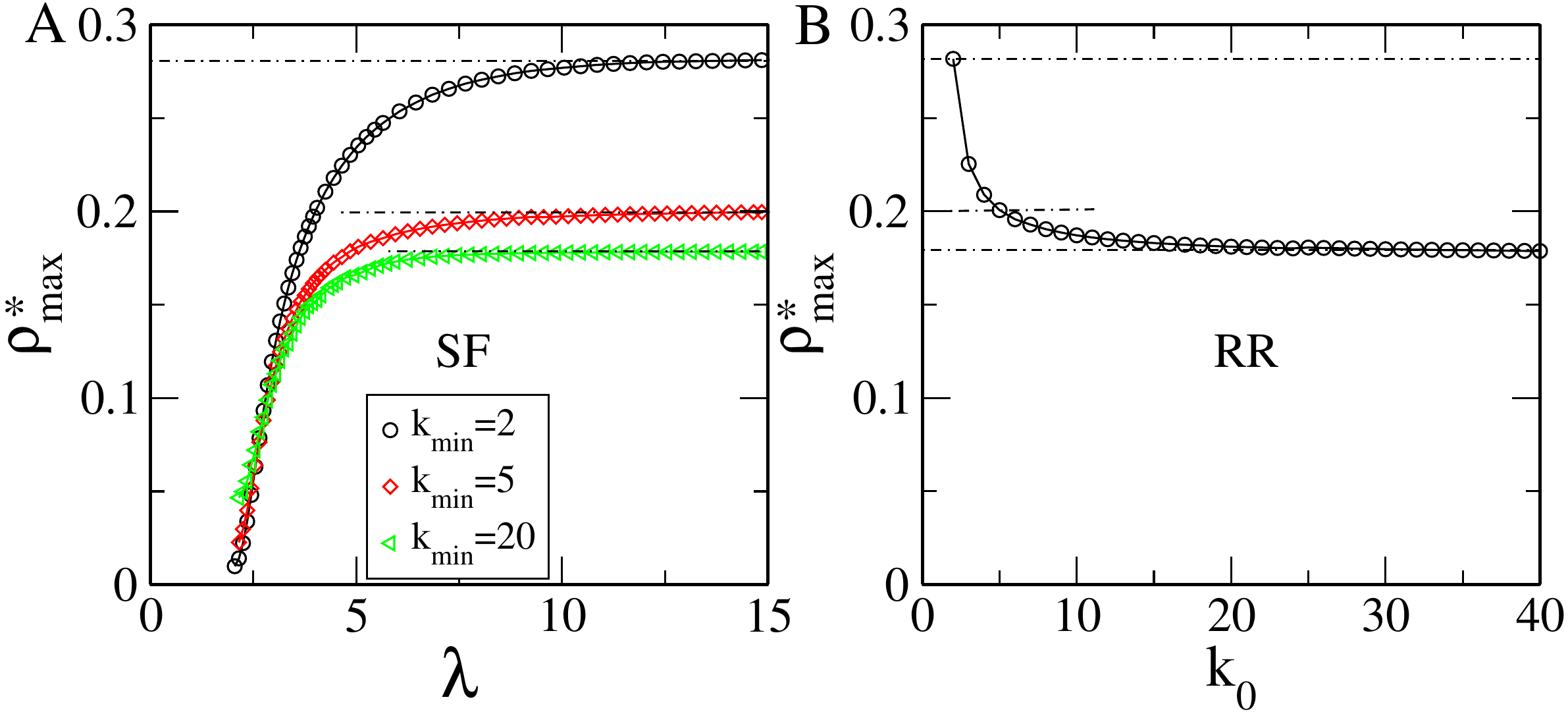}
\caption{(\textit{A}) $\rho^{\ast}_{\rm max}$ as a
  function of $\lambda$ for two fully interdependent SF networks with
  the same number of nodes and degree exponent and $k_{\rm min}=2$ (circle),
  5 (diamond) and 20 (triangle); $\rho^{\ast}_{\rm max}$ has an upper limit
  of $0.282$ (circle), $0.201$ (diamond) and $0.181$ (triangle) as
  $\lambda \rightarrow \infty$. (\textit{B}) $\rho^{\ast}_{\rm max}$ as a function of
  $k_0$ for two fully interdependent RR networks with the same number of
  nodes and $k_0$; $\rho^{\ast}_{\rm max}$ approaches $0.1756$ as
  $k_0\rightarrow \infty$.}\label{Fig5}  
\end{figure}

Similarly, we obtain $\rho^{\ast}_{\rm max}$ as a function of the degree
exponent $\lambda$ for two fully interdependent SF networks (see
Fig.~\ref{Fig5}\textit{A}) and $\rho^{\ast}_{\rm max}$ as a function of $k_0$ for
two fully interdependent RR networks (see Fig.~\ref{Fig5}\textit{B}).  Note
that as $\lambda$ increases, $\rho^{\ast}_{\rm max}$ initially increases but
later stabilizes at a value determined by $k_{\rm min}$ as the degree
distribution becomes more homogeneous and its network structure becomes
the same as that in an RR network with $k_0=k_{\rm min}$ (see \textit{SI, Appendix},
section 3.2).  For RR networks, as $k_0$ increases, $\rho^{\ast}_{\rm max}$
initially decreases but later stabilizes at a value close to 0.1756,
since at very large $k_0$ the structure of these RR networks resembles
that of ER networks with $\left\langle k\right\rangle=k_0$ (see
\textit{SI Appendix}, section 3.2).

Next we solve $\rho^{\ast}_{\rm max}$ of two fully interdependent networks
as a function of $\alpha$, where $\alpha={\left\langle k
  \right\rangle_A}/{\left\langle k \right\rangle_B}$ (see Fig. S10 in
\textit{SI Appendix}, section 4.1).  We find that in two ER networks, as
$\alpha$ increases, $\rho^{\ast}_{\rm max}$ increases and has a maximum at
$\alpha = 1$, corresponding to the symmetric case studied above.  In the
case of RR networks with large $k_0$, $\rho^{\ast}_{\rm max}$ behaves
similarly to its counterpart in ER networks, peaking around $\alpha=1$
at 0.1756 (see  Fig.~\ref{Fig5}\textit{B}). 
Moreover, in the case of SF networks when $\lambda \in (2,3]$, $\rho^{\ast}_{\rm max} \leq 0.11$;
whereas when $\lambda$ and $k_{\rm min}$ are relatively large,
$\rho^{\ast}_{\rm max}$ will also peak around $\alpha=1$ with a value
close to that obtained in RR networks.  Therefore in the extreme case
where $\lambda$ and $k_{\rm min}$ are large, SF networks converge to RR
networks with $k_0=k_{\rm min}$, which further converge to ER networks
with $\left\langle k \right\rangle=k_0$.  Thus in these extreme cases
there exists a universal $\rho^{\ast}_{\rm max}$ equal to 0.1756 (see
\textit{SI Appendix}, section 4.2).

Our approach can be generalized to solve the case of tree-like networks of
networks (NON) \cite{gao2012networks,Bianconi2014PRE}. For example, we
study the symmetric case of an ER NON with $n$ fully interdependent
member networks and obtain
\begin{equation}
\rho^{\ast}_{\rm max}=1-e^{1-1/n}/n,
\end{equation}
which is independent of the average degree $\left\langle k
\right\rangle$ (see \textit{SI Appendix}, section 3.1.2).  This relationship
indicates that the bigger $n$ is, the larger $\rho^{\ast}_{\rm max}$ should
be, which is consistent with the previous finding that the more networks
an NON has, the more vulnerable it will be \cite{gao2012networks}.

\section{Test on Empirical Data}
\begin{figure}[h!]
\centering
   \includegraphics[width=0.45\textwidth, angle = 0]
   {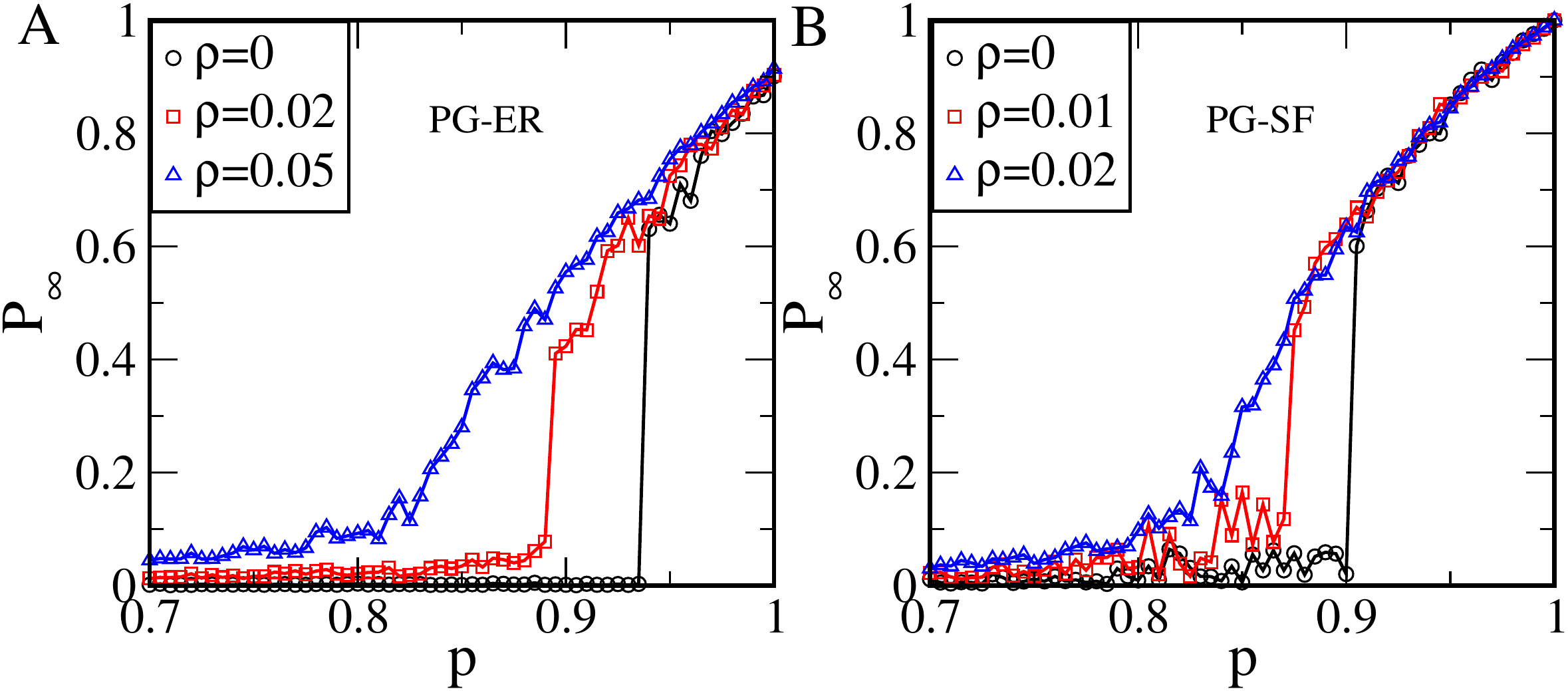}
\caption{ Percolation transition in real-world
  systems with the introduction of reinforced nodes. (\textit{A}) The circles,
  squares and triangles represent simulation results of a system
  composed of the US power grid (PG, with $N=4941$, $\left\langle k
  \right\rangle=2.699$) and an ER network ($N=4941$, $\left\langle k
  \right\rangle=2.699$) with interdependence strength $q=0.65$ and
  $\rho=0,0.02,0.05$ respectively; (\textit{B}) The circles, squares and
  triangles represent simulation results of a system composed of the
  same PG and an SF network ($N=4941$, $\lambda=2.7$, $k_{\rm min}=2$) with
  interdependence $q=0.65$ and $\rho=0,0.01,0.02$ respectively. The
  symbols are results obtained from a single realization.}\label{Fig6}
\end{figure}

We next test our mathematical framework on an empirical network, the US power grid
(PG) \cite{watts1998collective}, with the introduction of a small
fraction of reinforced nodes. It is difficult to establish the exact
structure of the network, that PG interacts with, and their interdependencies due 
to lack of data. However, to get qualitative insight  into the problem we 
couple the PG with either ER or SF networks
which can be regarded as approximations to many real-world networks. 
Our motivation is to test how our model performs in the interdependent 
networks system with some real-world network features. Note that here, 
our results present cascading failures due to structural failures and do not
 represent failures due to real dynamics, such as cascading failures due to 
 overloads, that appear in power grid network system.
Figure~\ref{Fig6} compares the mutual percolation of two systems of
interdependent networks with the same interdependence strength: PG
coupled to a same sized ER network (Fig.~\ref{Fig6}\textit{A}) and PG coupled
to a same sized SF network (Fig.~\ref{Fig6}\textit{B}).  As discussed above,
for $\rho$ below a certain critical value $\rho^{\ast}$ the systems will
undergo abrupt transitions, whereas for $\rho$ above $\rho^{\ast}$ the
systems do not undergo any transition at all.  We also find that, for
the interdependence strength $q=0.65$ shown here, the $\rho^{\ast}$
value of the latter case is very small and close to 0.02 (Fig.~\ref{Fig6}\textit{B}).
 
 \section{Summary}
In summary, we have developed a general percolation framework for studying 
interdependent networks by introducing a fraction of reinforced nodes at random. 
We show that the introduction of a relatively small fraction of reinforced nodes,
$\rho^{\ast}$, can avoid abrupt collapse and thus enhance
its robustness.  By comparing $\rho^{\ast}$ in ER, SF and RR networks,
we reveal the close relationship between these snetwork structures of these
networks in extreme cases and find the universal upper bound for
$\rho^{\ast}$ to be 0.1756.  We also observe improved robustness in
systems with some real-world network structure features. The
framework presented here might offer some useful suggestions on how
to design robust interdependent networks.

\begin{acknowledgments}
We wish to thank ONR (Grant N00014-09-1-0380, Grant N00014-12-1-0548, Grant N62909-14-1-N019),
DTRA (Grant HDTRA-1-10-1-0014, Grant HDTRA-1-09-1-0035), NSF (Grant CMMI 1125290),
 the European MULTIPLEX, CONGAS and LINC projects, the Next Generation Infrastructure (Bsik),
 and the Israel Science Foundation for financial support. We also thank the FOC program 
 of the European Union for support. YH is supported by the NSFC grant NO. 61203156.
\end{acknowledgments}


\end{document}